%% file: main.tex
\documentclass[twocolumn]{aastex701}
\usepackage{apjfonts}
\usepackage{rotating}
\usepackage{amsmath, color,graphicx,tablefootnote,subfigure}
\usepackage{graphics,subfigure,hyperref,float}
\usepackage[rightcaption]{sidecap}

\newcommand{\D}{$^{\dagger}$}

\newcommand{\W}{$\lambda$}

\newcommand{\HII}{H$\,${\sc ii}}
\def\hii  {\ifmmode{{\rm H}{\rm \small II}}\else{H\ts {\scriptsize II}}\fi}

\newcommand{\Yp}{$Y_{\rm p}$}

\def\beq{\begin{equation}}
\def\eeq{\end{equation}}
\def\alwaysmath#1{{\ifmmode{#1}\else{$#1$}\fi}}
\def\he#1{\hbox{\alwaysmath{{}^{#1}}{\rm He}}}
\def\li#1{\hbox{\alwaysmath{{}^{#1}}{\rm Li}}}
\def\sun{${\,_\odot}$}
\def\10830{{He~I $\lambda$10830}}
\def\3889{{He~I $\lambda$3889}}

\shorttitle{The LBT $Y_{\rm p}$ Project I}
\shortauthors{Skillman et al.}

\begin{document}

\title{The LBT $Y_{\rm p}$ Project I: An Improved Determination of the Primordial Helium Abundance -- Project Description, Sample Selection, Observations, and Methodology}


\author[0000-0003-0605-8732]{Evan D.\ Skillman}
\affiliation{Minnesota Institute for Astrophysics, University of Minnesota, 116 Church St. SE, Minneapolis, MN 55455}
\email{skill001@umn.edu}
\author[0000-0003-1435-3053]{Richard W.\ Pogge}
\affiliation{Department of Astronomy, The Ohio State University, 140 W 18th Ave., Columbus, OH, 43210}
\affiliation{Center for Cosmology \& AstroParticle Physics, The Ohio State University, 191 West Woodruff Avenue, Columbus, OH 43210}
\email{pogge.1@osu.edu}
\author[0009-0006-2077-2552]{Erik Aver}
\affiliation{Department of Physics, Gonzaga University, 502 E Boone Ave., Spokane, WA, 99258}
\email{aver@gonzaga.edu}
\author[0000-0002-0361-8223]{Noah S.\ J.\ Rogers}
\affiliation{Center for Interdisciplinary Exploration and Research in Astrophysics (CIERA), Northwestern University, 1800 Sherman Avenue, Evanston, IL 60201, USA}
\email{noah.rogers@northwestern.edu}
\author[0000-0003-4912-5157]{Miqaela K.\ Weller}
\affiliation{Department of Astronomy, The Ohio State University, 140 W 18th Ave., Columbus, OH, 43210}
\email{weller.133@buckeyemail.osu.edu}
\author[0000-0001-7201-5998]{Keith A. Olive}
\affiliation{William I. Fine Theoretical Physics Institute, School of Physics and Astronomy, University of Minnesota, Minneapolis, MN 55455, USA}
\email{olive@umn.edu}
\author[0000-0002-4153-053X]{Danielle A. Berg}
\affiliation{Department of Astronomy, The University of Texas at Austin, 2515 Speedway, Stop C1400, Austin, TX 78712, USA}
\email{daberg@austin.utexas.edu}
\author[0000-0001-8483-603X]{John J. Salzer}
\affiliation{Department of Astronomy, Indiana University, 727 East Third Street, Bloomington, IN 47405, USA}
\email{josalzer@iu.edu}
\author[0000-0002-2901-5260]{John H. Miller, Jr.}
\affiliation{Minnesota Institute for Astrophysics, University of Minnesota, 116 Church St. SE, Minneapolis, MN 55455}
\email{mill9614@umn.edu}
\author[0009-0009-2024-9317]{Jayde Spiegel}
\affiliation{Department of Astronomy, The Ohio State University, 140 W 18th Ave., Columbus, OH, 43210}
\affiliation{Department of Physics, University of California Santa Cruz, 1156 High Street, Santa Cruz, CA 95064}
\email{jdspiegel1221@gmail.com}
\author[0000-0002-4137-5306]{Tsung-Han Yeh}
\affiliation{TRIUMF, 4004 Wesbrook Mall, Vancouver, BC V6T 2A3, Canada}
\email{thyeh@triumf.ca}
\author[0000-0002-4188-7141]{Brian D. Fields}
\affiliation{Department of Astronomy, University of Illinois, Urbana, IL 61801}
\affiliation{Department of Physics, University of Illinois, Urbana IL 61801}
\affiliation{Illinois Center for Advanced Studies of the Universe}
\email{bdfields@illinois.edu}

\begin{abstract}

Extremely low metallicity \HII{} regions have been observed with the goal of determining the primordial helium abundance ($Y_{\rm p}$).  
$Y_{\rm p}$, combined with standard big bang nucleosynthesis and the half-life of the neutron, provides a direct measurement of the number of neutrino families, but $Y_{\rm p}$ must be measured very precisely to provide meaningful constraints on physics beyond the Standard Model. 
Here we describe a program to combine new Large Binocular Telescope (LBT) observations with a new analysis methodology to significantly improve the determination of $Y_{\rm p}$. 
The LBT, with its MODS and LUCI instruments, produces spectra, which, when combined  with our new analysis methodology, are capable of delivering He abundances in individual \HII{} regions with uncertainties of approximately 2\% or less.  
Archival LBT/MODS spectra of standard stars over a four-year period enable the determination of a wavelength-dependent uncertainty in the MODS spectral response, resulting in improved relative emission line uncertainties. 
An optimized sample of low-metallicity galaxies has been selected with the goal of producing a determination of $Y_{\rm p}$ with a precision of $\sim$ 0.5\%, sufficient to provide an independent constraint on the effective number of neutrino families of $\sim$ 3\%.

\end{abstract}



\section{Introduction}\label{sec1}

The LBT \Yp\ Project is designed to produce a robust, high
precision measurement of the primordial helium abundance (\Yp ).

\subsection{Project Motivation}\label{motivation}

Standard big bang nucleosynthesis \citep[SBBN;][]{olive2000, Iocco:2008va,cyburt2016, Pitrou:2018cgg, fields2020} remains one of the deepest available probes to the early universe. 
It is also an important probe of physics beyond the Standard Model \citep{Sarkar:1995dd,Cyburt2005,Yeh:2022heq,Jedamzik:2009uy,Pospelov:2010hj}. 
New physics, which affects the idealized conditions of a radiation-dominated thermal bath at a temperature of roughly 1 MeV, has the potential of breaking the agreement between theory and the observational determinations of the $\he4$ and deuterium abundances. For example, theories which include new particle degrees of freedom lead to an increased rate of expansion during nucleosynthesis, leaving less time for the conversion of neutrons to protons, and hence lead to an increased helium mass fraction \citep{Steigman:1977kc,cyburt2016, fields2020,Yeh:2022heq}.
These new particle degrees of freedom are often scaled as contributions to the number of very light neutrinos, $N_\nu$. 
The upper limits on $N_\nu$ from BBN rely heavily on accurate measurements of the primordial helium abundance, $Y_{\rm p}$.

Although the cosmic microwave background (CMB) measurements  by {\it WMAP} \citep{komatsu2014} and {\it Planck} \citep{aghanim2020} provide an unparalleled determination of the fundamental parameters of the standard cosmological model, including cold dark matter and dark energy ($\Lambda$CDM), big bang nucleosynthesis and the observations of the light element abundances also offer an important cross-check, in particular on the determination of the baryon density.  The most recent {\it Planck} result for the baryon density  (which takes into account the correlation between the baryon density and the helium abundance), $\Omega_B h^2 = 0.02237 \pm 0.00015$, corresponds to a baryon-to-photon ratio of $\eta = (6.12 \pm 0.04) \times 10^{-10}$.  Because the uncertainty in $\eta$ is now less than 1\%, SBBN (defined with $N_\nu = 3$) has effectively become a parameter-free theory \citep{cyburt2002}, and relatively precise predictions of the primordial abundances of the light elements D, $^{3}$He, $^{4}$He, and $^{7}$Li are available \citep[e.g.,][]{fields2020}.  
While the \li7 abundance remains problematic \citep[e.g.,][]{Cyburt:2008kw,Fields:2022mpw}, D/H determinations from quasar absorption systems have become quite precise, and are in excellent agreement with the prediction from SBBN and the CMB \citep[e.g.,][]{cooke2014, cooke2016, cooke2018, riemer2015, riemer2017, balashev2016,Zavarygin:2018,guarneri2024MNRAS.529..839G,kislitsyn2024MNRAS.528.4068K}.

Using a neutron mean life of $878.4 \pm 0.5$ s \citep{rpp} 
and the {\it Planck} value of $\eta$, SBBN yields a primordial abundance for $^{4}$He, $Y_{\rm p}$ $= 0.2467 \pm 0.0002$ \citep{Yeh:2022heq,Yeh:2023nve}. 
By allowing $Y_{\rm p}$ to vary as an independent parameter, fits to CMB anisotropies allow for a determination of $Y_{\rm p}$ within the context of $\Lambda$CDM.  
The recent {\it Planck} results found $Y_{\rm p}$ $= 0.239 \pm 0.013$ (68\% CL) \citep{aghanim2020}.  This 5.4\% uncertainty on $Y_{\rm p}$, from CMB measurements alone, is currently not competitive with other constraints; the helium abundance from emission line measurements from metal-poor galaxies provides significantly better precision, cf. $Y_{\rm p}$ = $0.2448 \pm 0.0033$ \citep{aver2022}, a 1.3\% determination.

More recently, the Atacama Cosmology Telescope (ACT) has combined ground based measurement of the CMB at high multipoles with {\it Planck} data and Dark Energy Survey Instrument (DESI) Baryon Acoustic Oscillations (BAO) data to obtain $Y_{\rm p}$ $= 0.2312 \pm 0.0092$ (68\% CL) - a 4\% uncertainty \citep{ACT:2025tim}. Similarly, the South Pole Telescope (SPT) has obtained $Y_{\rm p}$ $= 0.2285 \pm 0.0085$ (68\% CL) when SPT data is combined with {\it Planck} and ACT data \citep{SPT-3G:2025bzu}. This is a 3.7\% determination.  

A precise primordial abundance is available not only for \he4, but also for deuterium, which is measured via its Lyman lines in quasar absorption systems. The weighted mean over 12 systems gives D/H $= (2.513 \pm 0.028) \times 10^{-5}$. The SBBN prediction for D/H is $(2.506 \pm 0.083) \times 10^{-5}$ \citep{Yeh:2023nve}. For D/H, theory is lagging behind the precision of the observational data, primarily due to uncertainties in the nuclear cross sections for the $d(d,p)t$ and $d(d,n)\he3$ reactions. The agreement is nevertheless excellent.  

To test SBBN beyond D/H, it is clear that precise determinations of \he4 are necessary.
Since $Y_{\rm p}$ scales as $\eta^{0.04}$ in SBBN, it is unlikely that \he4 abundance measurements will ever be competitive with
the CMB and/or D/H determinations in fixing the baryon density since D/H scales as $\eta^{-1.634}$ in SBBN \citep{Yeh:2023nve}. Although D/H also scales more strongly with $N_\nu$ (as $N_\nu^{0.405}$) as compared with \he4 (which scales as $N_\nu^{0.163}$), due to the uncertainties in the BBN predictions of D/H, $Y_{\rm p}$ provides the strongest constraints available on $N_\nu$ and hence on the physics of the early universe beyond the Standard Model \citep{Cyburt2005,Yeh:2022heq}.  
Currently, the combination of {\it Planck} CMB data with abundance measurements used in BBN has a maximum likelihood with $N_\nu = 2.898 \pm 0.141$, just below the Standard Model value of 3; the uncertainty in this value depends sensitively on the data considered \citep{fields2020,Yeh:2022heq}.
For example, the CMB alone can fix $N_\nu$ only with a precision of $\pm 0.29$ (68\% CL). However, when {\it Planck} data are combined with the BBN relation between $Y_{\rm p}$ and $\eta$, this uncertainty improves to $\pm 0.19$ without using any abundance data. Current helium abundance data further improve the uncertainty to $\pm 0.15$, while D/H (without $Y_{\rm p}$) gives $\pm 0.18$ and is competitive with \he4. 
 
More recent CMB experiments find increased precision for the effective number of neutrinos at later epochs. In the Standard Model, $N_{\rm eff} = 3.044$ \citep{Drewes:2024wbw} \footnote{$N_{\rm eff}$, the effective number of neutrino flavors, is a measure of the total energy density of relativistic particles in the early universe. The Standard Model value for $N_{\rm eff}$ is slightly greater than the 3 Standard Model neutrino flavors due to the slightly higher temperature from e$^+$e$^-$ annihilation before neutrinos are completely decoupled.} and does not evolve between BBN and CMB decoupling. For example,  ACT combined with {\it Planck} and DESI BAO obtain $N_{\rm eff} = 2.86 \pm 0.13$ \citep{ACT:2025tim}.
SPT data combined with {\it Planck} and ACT give $N_{\rm eff} = 2.81 \pm 0.12$ \citep{SPT-3G:2025bzu}. 
Indeed, it is hoped that future CMB missions can improve these uncertainties $< \pm 0.09$ for the CMB alone \citep{abazajian2019}, and $< \pm 0.06$ when the CMB is combined with BBN and $Y_{\rm p}$ determinations \citep{Yeh:2022heq}. 
The primary goal of the LBT \Yp\ Project is to reduce the
uncertainty on $Y_{\rm p}$ from 1.3\% to $\sim$ 0.5\%.  This reduction in the uncertainty in $Y_{\rm p}$ translates into
an uncertainty on the number of neutrino families from 0.14 to $\sim$0.08.

An improvement in the uncertainty in $N_\nu$ of this magnitude will allow one to probe in detail the decoupling period prior to BBN.  
Ultimately, the goal of distinguishing between the Standard Model number of light neutrinos $N_\nu = 3$ and the effective number of neutrinos affecting the expansion rate of the Universe, $N_{\rm eff} = 3.044$, seems within reach. 
Furthermore, barring a systematic shift in the maximum likelihood value for $N_\nu$, the reduction in its uncertainty, or, equivalently, the uncertainty on $N_{\rm eff}$, will place unprecedented constraints on physics beyond the Standard Model. Physics beyond the Standard Model may also induce a change in the effective number of degrees of freedom between BBN and CMB decoupling. This too, is an interesting challenge, perhaps also within reach.  

As noted above, certain physics beyond the Standard model can affect the expansion rate of the Universe at the time of BBN. The expansion rate, characterized by the temperature-dependent Hubble parameter is primarily determined by the energy density in radiation (at the time of BBN),
\beq
H^2(T) \simeq \frac{8\pi}{3}G_N \rho_{\rm R} \, ,
\eeq
where $G_N$ is Newton's constant and, around the time of BBN, 
\beq
\rho_{\rm R} = \rho_\gamma + \rho_e + \rho_\nu = \frac{\pi^2}{30} \left(2+\frac72+\frac74N_\nu \right)T^4 \, .
\eeq
Thus, new physics contributing to the energy density can be constrained by its effective contribution to $N_\nu$. For example, in models of neutrino masses which involve light right-handed or sterile neutrinos, the interactions of these states must be weak enough so that they decouple long before Standard Model neutrino decoupling at $\simeq 2$~MeV, suppressing their temperature relative to the left-handed neutrinos and hence their contribution to $N_\nu$. This may set lower limits to the mass of an extra gauge boson mediating right-handed interactions (see e.g. \citet{Steigman:1986nh,Gonzalez-Garcia:1989ygi,Barger:2003zh}) or the degree of mixing between sterile neutrinos and left-handed neutrinos \citep{Dolgov:2003sg,boser2020,Alonso-Alvarez:2022uxp}.  Similar limits would apply to any dark radiation present at the time of BBN or to a non-relativistic matter component \citep{sabti2020,Yeh:2024ors}. Other limits include constraints on primordial gravitational waves \citep{Boyle:2007zx,kohri2018}. 

Determinations of $Y_{\rm p}$ are also relevant to attempts to resolve the Hubble tension \citep[e.g.,][]{riess2019, riess2021}.
In particular, BAO and BBN data (via primordial deuterium and helium abundances) allow for a CMB-independent determination of H$_0$ for comparison with distance ladder determinations \citep{addison2018, schoneberg2019}. 
Furthermore, many models of new physics beyond the Standard Model proposed to resolve the Hubble tension often require increases in $N_{\rm eff}$ or are motivated in their design by limits on it. Correspondingly, limits on $N_{\rm eff}$ inferred from measurements of $Y_{\rm p}$ help constrain models of new physics proposed for resolving the Hubble tension. 
Prominent, outstanding questions in cosmology, such as the Hubble tension, and the range of extensions beyond the Standard Model which are constrained by BBN and the measurement of $Y_{\rm p}$, demonstrate the importance of a better, higher precision determination of $Y_{\rm p}$.

\subsection{Present State of the Art}

\subsubsection{Recent Methodology Improvements}

The most precise values of $Y_{\rm p}$ are determined by fitting the helium abundance versus a measurement of metallicity (e.g., oxygen) from \HII\ regions, and extrapolating back to very low metallicity as first demonstrated by \citet{peimbert1974}.  
Thus, improvements in the uncertainty on the extrapolation can come from reducing the uncertainties on individual measurements and by increasing the number of high-quality measurements.   
Historically, obtaining better than $\sim$ 3\% precision for individual objects has been a challenge.  
\he4 abundance determinations are vulnerable to systematic uncertainties and degeneracies among the input parameters needed to model emission line fluxes \citep{olive2001, olive2004, peimbert2003, peimbert2007, izotov2007, aver2010}.
Using Monte Carlo methods reveals these degeneracies, puts these determinations on a firm statistical basis, and allows an objective comparison between data and theoretical models \citep{aver2011, aver2012}.
Curiously, \citet{aver2012} discovered that, in many cases, the models used to extract abundances are statistically inconsistent with the observations (in that those solutions fail to pass a standard 95\% CL $\chi^2$ test) and employed corresponding quality cuts on their dataset.  
Subsequent analyses by other groups adopting the \citet{aver2012} methodology found similar inconsistencies \citep{fernandez2019, hsyu2020}, and others have employed stringent cuts on their observational data for related reliability concerns \citep{izotov2013, izotov2014}.
Our recent research is consistent with the hypothesis that these incompatibilities are primarily due to deficiencies in the observations.  In particular, note that many of these observations are now quite old -- the majority of the analyzed targets come from the HeBCD dataset, with observations dating back to the 1990s \citep{izotov2007}.  
Using higher quality, higher spectral resolution, broader wavelength coverage observations has generally produced statistically consistent model solutions. 

For example, \citet{aver2012} started with a dataset of 93 observations of 86 \HII\ regions in 77 galaxies from \citet{izotov2007}. It was argued that the absence of He~I $\lambda$4026 in 23 of the observations led to a systematic bias towards higher \he4, and those observations without $\lambda$4026 were discarded. However, analysis of the remaining 70 observations, found that, in 45 of them, the comparison between model and data resulted in $\chi^2 > 4$, thus indicating a poor model fit to the data at the 95.45\% CL, given nine observed line ratios and eight model parameters used in that model.
Further cuts were made due to solutions with very high ($>25\%$) neutral hydrogen fraction, or very high oxygen abundances, leaving 22 observations, of which eight were flagged with either a large optical depth or large underlying absorption in either H or He. 
Thus, the vast majority of the observations were deemed unsuitable for further analysis. 
In fact, when \citet{aver2013} used improved emissivities from \citet{porter2012, porter2013} to calculate a revised value of $Y_{\rm p}$, the solutions for only two objects improved to an acceptable level giving a total of 16 observations to deduce $Y_{\rm p}$, indicating that inaccuracies in emissivities were likely {\it not} the dominant cause of statistically inconsistent solutions for targets.

A significant breakthrough came with the addition of observations of the NIR \10830 emission line, as first suggested by \citet{izotov2014}, which led to significant improvements in the \he4 abundance determinations \citep{izotov2014, aver2015}.
The \10830 line is particularly useful as it helps break some of the degeneracies in the parameters used to derive the \he4 abundance, most notably that between the electron density, $n_e$, and temperature, as highlighted by \citet{aver2010}.
For example, 31 targets from \citet{izotov2007} contained \10830 observations \citep{izotov2014}, and these were subject to the MCMC analysis now with ten flux ratios to fit the same eight parameters. 
Without \10830, only 11 of the 31 objects survived the various cuts employed. With \10830, that total grew to 16 (though two corresponded to independent observations of the same galaxy). 
This result strongly highlights the benefits of expanding the model to employ additional observations.

In a study of the extremely metal poor object Leo~P, \citet{aver2021} demonstrated that taking advantage of the large wavelength baseline of LBT/MODS enables the use of additional hydrogen and helium emission lines and results in better constrained model parameters, particularly the reddening.
Additional improvements were made to the treatment of the collisional excitation of H~I.  After the required addition of collisional excitation rates to higher levels, adding more hydrogen emission lines provided a better constraint on the impact of the collisional excitation.  The correction for underlying stellar absorption was revisited, and through the use of BPASS stellar evolution and spectral synthesis models of \citet{eldr2009, stan2018}, a uniform set of scaling relations was developed to include all of the newly added hydrogen and helium emission lines.
\citet{aver2021} also revisited the procedure for handling the blended line \3889 with H8, including its underlying absorption and radiative transfer.

\citet{kurichin2021b, kurichin2024, kurichin2025} represent a series of suggested improvements in the methodology for deriving $Y_{\rm p}$.  \citet{kurichin2021b} addresses the often used method of deriving the value of O/H when only one measurement of temperature is available.  Typically, the temperature in the high ionization zone derived from \ion{O}{3} observations is used to infer a temperature in the low ionization zone, and \citet{kurichin2021b} point out that the inferred low ionization zone temperature can be biased and its uncertainty can be underestimated \citep[as noted previously in other contexts by, e.g.,][]{arel2020, roge2021, roge2022}. As a result,  O$^+$/H$^+$ can be biased, resulting in a biased measurement of O/H. In the LBT $Y_{\rm p}$ project, we obtain multiple direct temperature measurements which allow us to make a better estimate of the appropriate temperature for the low ionization zone and its uncertainty.

\citet{kurichin2024} present a suggested improvement for the treatment of the underlying absorption in the \ion{H}{1} and \ion{He}{1} lines.  In brief, they model the underlying absorption based on full SED-fitting of the observed galaxy spectrum, including stellar and nebular continua as well as emission-line profiles. This is claimed to be superior to the methodology of, e.g., \citet{aver2021}, which scales the equivalent widths of individual lines based on numerical studies of the emission lines in single stellar populations \citep[e.g.,][]{eldr2017, stan2018}.  We have not been able to test and compare these techniques, but we do note that our modeling of underlying absorption produces very reasonable minimization values.

\citet{kurichin2025} produce a new grid of \ion{He}{1} radiative transfer corrections.  This grid is consistent with modern \ion{He}{1} emissivities and identifies flaws in in the previously used corrections \citep{benjamin2002}.  We greatly appreciate this new contribution and we have adopted these radiative transfer corrections for the LBT $Y_{\rm p}$ project.

\begin{figure}[t!]
\resizebox{\columnwidth}{!}{\includegraphics{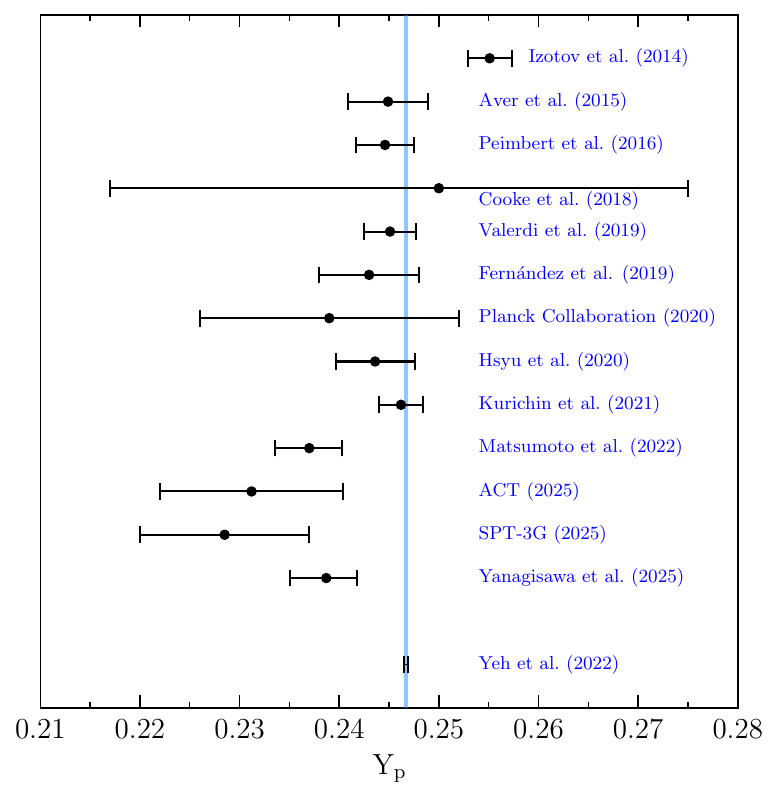}}
\caption{
A comparison of recent $Y_{\rm p}$ results.  The SBBN result based on the Planck determined baryon density \citet{aghanim2020} is plotted as a vertical band in light blue \citep{Yeh:2022heq,Yeh:2023nve}.  
}
\label{figure:Yp_Comp}
\end{figure}

\subsubsection{Recent Results} \label{Sec:RecentResults}

Fig.~\ref{figure:Yp_Comp} gives a visual representation of recent determinations of \Yp, listed in chronological order and compared to the SBBN result based on the Planck determined baryon density \citep{Yeh:2022heq}.  All of the points were determined through observations of \HII\ regions in metal-poor galaxies except for the \citet{cook2018b} point, from observations of an absorption system, and three determinations based on CMB observations \citep{aghanim2020, ACT:2025tim, SPT-3G:2025bzu} as described earlier in Section~\ref{motivation}.  In this subsection, we will described the determinations based on \HII\ region observations in more detail.

\citet{aver2015} derived a value of $Y_{\rm p}$ $=$ 0.2449 $\pm$ 0.0040 from a sample of 15 objects which passed the $\chi$$^2$ cut-off.  This value was in tension with with value of $Y_{\rm p}$ $=$ 0.2551 $\pm$ 0.0022 previously derived by \citet{izotov2014}.  While both analyses were drawn from the same observational database, the analysis methodologies and samples were different.  \citet{izotov2014} emphasized that their result corresponded to a value of $N_{\rm eff}$ $=$ 3.58 $\pm$ 0.25, implying a significant departure from the Standard Model.   The value from \citet{aver2015} is consistent with SBBN.

These studies were closely followed by others.
\citet{peimbert:2016} used observations of five objects and an assumed slope of $\Delta Y/\Delta {\rm O} = 3.3 \pm 0.7$ to derive $Y_{\rm p}$ $= 0.2446 \pm 0.0029$. 
In a similar approach, \citet{valerdi2019} combined a very precise measurement of He/H in the Small Magellanic Cloud \HII\ region NGC~346 with the same assumed slope of $\Delta Y/\Delta {\rm O}$ to obtain a value of $Y_{\rm p}$ $= 0.2451 \pm 0.0026$.
The more recent analysis of \citet{fernandez2018}, using SDSS-III data with regressions in O/H, N/H, and S/H, found $Y_{\rm p}$ $= 0.245 \pm 0.007$, which improved to $Y_{\rm p}$ $= 0.243 \pm 0.005$, when Bayesian statistical techniques were applied \citep{fernandez2019}.
\citet{fernandez2018} use the collisionally excited [S~III] lines to derive a nebular temperature, and this appears to be a good choice for \HII\ regions where the quality of the spectra does not allow a measurement of the temperature from the He~I lines (as opposed to a temperature measurement from the [O~III] lines which may be more strongly biased toward higher temperatures).
All four of these independent analyses agree with the result of \citet{aver2015} within the limits of their uncertainties.

\citet{hsyu2020} presented new Keck NIRSPEC and NIRES Keck observations of \10830 in 16 galaxies identified from SDSS imaging and having existing optical spectra \citep{hsyu2018}.
Their analysis followed the \citet{aver2015} model, and, in their primary data sample, they combined their new observations with 13 targets from the HeBCD sample and 38 targets from the Sloan Digital Sky Survey (SDSS).  From the combined dataset they obtained $Y_{\rm p}$ $= 0.2436^{+0.0039}_{-0.0040}$ in good agreement with \citet{aver2015}.
Importantly, \citet{hsyu2020} highlight the fact that only a small fraction of their newly observed targets (3 targets in their primary sample) passed their quality cuts for successful modeling and noted that their highest quality spectra were the most likely to pass.  While \citet{hsyu2020} followed the \citet{aver2015} model, there are important differences that are discussed in Appendix~A.

This has been followed by \citet{Kurichin2021}, who employed the \citet{aver2015} model and methodology on 100 spectra from the SDSS, combined with 20 spectra from the HeBCD database. Their resulting measurement of $Y_{\rm p}$ is consistent with the results of \citet{aver2015} and \citet{hsyu2020}, but with an uncertainty reduced by almost a factor of two.  We have some concerns about the \citet{Kurichin2021} sample since it is dominated by SDSS spectra, which tend to be quite low SNR.  Furthermore, the SDSS spectra used by \citet{hsyu2020} and \citet{Kurichin2021} are dominantly at higher metallicities (O/H $\ge$ 8 $\times$ 10$^{-4}$) and generally have larger uncertainties (typically larger than 5\%, in part, reflecting the absence of \10830 measurements).  Nonetheless, the uncertainty on $Y_{\rm p}$ is reduced simply due to the large number of targets. 

It is our opinion that adding large numbers of low quality points, while statistically effective, is not the best path forward.
In a discussion of how best to improve the measurement of $Y_{\rm p}$, \citet{ferland2010} provide several arguments as to why it is better to use exclusively high quality spectra. 
The SDSS spectra reduce the uncertainty on $Y_{\rm p}$ primarily by reducing the uncertainty on the slope of the Y versus O/H relationship.  This increases the vulnerability to the assumption of a perfectly linear relationship, which is not necessarily to be expected given the different nucleosynthesic origins of helium and oxygen (or nitrogen or sulfur).   Additionally, at high metallicity, our methodology cannot strongly constrain the neutral hydrogen fraction (because it is determined through the collisional excitation of neutral hydrogen, which decreases precipitously with lower temperatures), although, at these low temperatures, this correction should be negligible.  Nonetheless, objects at higher metallicity have the potential for an additional systematic uncertainty.  Finally, the uncertainty in the mean can be underestimated if the dispersion is too large (see discussion in Paper IV).

Importantly, \citet{Kurichin2021} conducted a re-analysis of the \citet{izotov2014} study.  They identified three items in the \citet{izotov2014} analysis which they called into doubt: the dropping of $\lambda$ 3889 and $\lambda$ 7075; assuming a value for the \ion{He}{1} underlying absorption; and applying a hard prior in a limited range for the temperature based on T[O~III]. \citet{Kurichin2021} duplicated the results of \citet{izotov2014}, and then conducted a reanalysis without the three identified methodology concerns, and found a resulting value of $Y_{\rm p}$ in good agreement with SBBN.

\citet{matsumoto2022} added new \10830 observations of 10 extremely metal poor galaxies and used the \citet{hsyu2020} methodology to derive their helium abundances.  
Of these ten galaxies, the solutions from the MCMC code for five did not pass the $\chi$$^2$ criteria.  For the five that did qualify, intriguingly, the solutions for density, underlying stellar absorption equivalent width, and optical depth produced negative values (of parameters that are intrinsically positive).  The non-physical nature of these solutions were not addressed.
These five newly observed targets were added to 54 objects from \citet{hsyu2020} (only 8 of which have NIR spectroscopy) with metallicities reaching up to 40\% of the solar value to derive $Y_{\rm p}$ $=$ 0.2370$^{+0.0033}_{-0.0034}$. 
\citet{matsumoto2022} emphasize that their result is 1 $\sigma$ lower than the value of \citet{aver2015} and the following studies in agreement with that value.  
\citet{matsumoto2022} note that, in a reanalysis of the \citet{hsyu2020} observations, they obtain a nearly identical result and thus attribute the lower value of $Y_{\rm p}$ to the addition of the five newly observed extremely metal-poor (XMP)  galaxies.

This study has recently been followed up by \citet{yanagisawa2025}.  They add new NIR observations of 29 galaxies and combine these new observations with optical observations from the SDSS, Magellan, or literature sources. Of these 29, 15 galaxies do not pass a quality of fit criterion, 3 galaxies are dropped due to radiation softness, and 3 galaxies are dropped due to high N/O with a result that 10 pass their quality criteria.  These 10 are added to 58 galaxies from the \citet{matsumoto2022} study producing a $Y_{\rm p}$ slightly higher than that derived by \citet{matsumoto2022}, but within 1 $\sigma$.

There are clear differences in the methodologies, observations, and results of the various literature studies. Thus, it is highly desirable to address these potential systematic uncertainties with a large dataset, at low metallicities, that is uniformly observed and analyzed for the purpose of measuring \Yp.

\subsection{Overview of the Project Papers}

This project will be presented in six papers.  This paper (Paper~I) consists of introductory material (\S 1), descriptions of the sample and observations (\S 2, 3), a discussion of new H and He emissivities (\S 4), a description of the project database (\S 5), and a summary (\S 6).
Paper~II  \citep{Rogers2026} describes the new LBT MODS observations, the derived physical conditions, and the O/H abundances for the sample.
Paper~III \citep{Weller2026} describes the LBT LUCI near-IR observations and the (\ion{He}{1}\,$\lambda$10380)/(\ion{H}{1}\,P$\gamma$) ratio critical for constraining the density.
Paper~IV \citep{Aver2026} describes our improved methodology, the resulting He abundance measurements, and our new derivation of $Y_{\rm p}$.
Paper~V \citep{Yeh2026} presents calculations of the impact that our new derivation of $Y_{\rm p}$ has on the constraint on $N_\nu$ and, consequently, new physics.
Paper~VI (N.\ Rogers et al., in prep.) presents additional chemical abundance measurements, which are not directly involved in the determination of $Y_{\rm p}$, but are an additional important result of our high quality optical spectra.   Paper~VI also  provides access to all of the reduced data products.


\section{Sample Selection}

\subsection{New Discoveries of Low Metallicity Galaxies}

Beyond our demonstrated model improvements, we are currently statistics limited.
Clearly, enlarging the sample of XMP galaxies\footnote{Here we adopt the definition of an XMP galaxy as having 12 + log(O/H) $\le$ 7.35 \citep{guse2015,mcquinn2020}.
Note that not all of the targets in our sample are this metal-poor, but the most metal-poor targets are most desirable because they reduce the vulnerability to the assumption of a linear relationship between He/H and O/H.}, with well-determined He abundances, will lead to improved constraints on the value of $Y_{\rm p}$.
The XMP galaxies are the most relevant for measurements constraining $Y_{\rm p}$; however, historically, the searches for these galaxies have had very low yields \citep[see discussion in][]{sanchez2017ApJ}.
Fortunately, in recent years, there have been a number of discoveries of XMP galaxies 
\citep[e.g.,][]{hirschauer2016, yang2017, guseva2017, 
hsyu2017, hsyu2018, izotov2018a, izotov2019, senchyna2019,
pustilnik2020, pustilnik2021, kojima2020, kojima2021}.

Although many of these XMP galaxies are too low in emission line surface brightness to be viable for precise He abundance measurements, observations of the most promising of these galaxies, with quality comparable to the observations of Leo~P \citep{aver2021}, represents a very desirable goal.
In this regard, the DESI survey is especially promising.  From the early release observations, \citet{Zou2024} and \citet{Zinchenko2024} report numerous XMP galaxy discoveries, and recently \citet{Scholte2026} have produced a catalog of thousands of metal-poor star forming galaxies from the DESI Data Release 2.  Although many of these fail our sample selection criteria (see next subsection), there is potential to significantly improve the sample of suitable, lowest metallicity targets.

\subsection{The LBT $Y_{\rm p}$ Project Sample Criteria}

Over the course of this project we have established criteria which optimize the utility of various targets for use in determining $Y_{\rm p}$.  First and foremost, we prioritize the lowest metallicity targets as judged by measurements of their oxygen abundance.  With this choice, we are minimizing the impact of higher metallicity objects on our final determination of $Y_{\rm p}$.  Thus, we are prioritizing the measurement of $Y_{\rm p}$ over the measurement of $\Delta$Y/$\Delta$O.  This has not necessarily been the case in previous $Y_{\rm p}$ measurements.
For example, as an extreme case, \citet{valerdi2019} make a determination of $Y_{\rm p}$ using observations of a single HII region in the SMC and a choice for the value of $\Delta$Y/$\Delta$O.  It is our intention to have as little dependence on $\Delta$Y/$\Delta$O as possible.  This is primarily because, although all evidence to date supports it, assuming a strictly linear relationship between He/H and O/H (or N/H or S/H) implies an unquantifiable systematic uncertainty \citep{1993A&A...277...42P,Fields:1998gv}.  If our entire sample consisted of targets with sufficiently low O/H such that the uncertainty in $\Delta$Y/$\Delta$O were irrelevant, then a straightforward weighted average of the points, with no dependence on $\Delta$Y/$\Delta$O, would be the best possible path to a secure measurement of $Y_{\rm p}$.
Up to now, the number of XMP galaxies which meet our other criteria was not sufficient to put us in this regime. This work is an important step towards that goal.

Second, we prioritize sources with high emission-line fluxes.  In order to strongly constrain the parameters which determine He/H (e.g., temperature, density, reddening, optical depth, underlying absorption, neutral fraction), we require high S/N in many intrinsically weak emission lines.  In the end, we are seeking uncertainties in the helium abundances for individual targets of $\sim$ 3\% or less.  Thus, high fluxes are a requirement.  Experimentally, we have determined that targets with H$\beta$ fluxes above 10$^{-15}$ erg s$^{-1}$ cm$^{-2}$  are required. For example, the Leoncino dwarf galaxy \citep{aver2022} with an H$\beta$ flux of 9 $\times$ 10$^{-16}$ erg s$^{-1}$ cm$^{-2}$ does not produce uncertainties on He/H as low as we desire (although we retain it in our sample).  On the other hand, Leo~P \citep{aver2021}, with an H$\beta$ flux of 3 $\times$ 10$^{-15}$ erg s$^{-1}$ cm$^{-2}$ achieves a desirable uncertainty.

Finally, we prioritize sources with high emission line equivalent widths and have determined a guideline in H$\beta$ equivalent width.  Even if a target has a sufficiently high H$\beta$ flux, if the continuum is strong, the weak emission lines will be lost to the random fluctuations in the continuum.  Experimentally, we have determined that a target requires an H$\beta$ equivalent width $\ge$ 100 \AA.  The F(H$\beta$) and EW(H$\beta$) goals developed for the project were based on our experience with the accumulating LBT observations, but since the subsequent targets were chosen based on literature values, not all of the observed targets met these goals.  It follows that those at higher metallicity or having less constrained values of He/H are less useful to the determination of $Y_{\rm p}$.

A complete listing of our LBT observations is given in Table~\ref{t:observations}. This table lists the target name, coordinates, redshift, and the MODS and LUCI integration times.  The tabulated integration times represent the total observing time for one LBT instrument, i.e., if a binocular observation lasted one hour, the table gives an integration time of 120 minutes. 
For the targets without LUCI integration times, it was determined from some feature(s) of their MODS observations that these are not viable targets. The final column provides notes, e.g., alternate target names, whether the MODS spectrum came from the archive, if the target is not viable  Table~\ref{t:observations} therefore lists all of the observations, some of which are not used in the final \Yp\ determination. The complete final sample is presented in paper IV.

\input{ObservationsTable}


\section{Observations}

\subsection{Optical Spectroscopy}

The LBT $Y_{\rm p}$ project utilizes the Multi-Object Double Spectrographs \citep[MODS,][]{pogg2010} on the Large Binocular Telescope \citep[LBT,][]{hill2010} to obtain the optical spectra of \ion{H}{2} regions in metal-poor galaxies. The MODS blue channel has a wavelength coverage of 3200--5700 \AA\ and R$\sim$1850 for the G400L (400\,lines\,mm$^{-1}$) grating; the red channel has a wavelength range of 5500--10000 \AA\ and R$\sim$2300 for the G670L (250\,lines\,mm$^{-1}$) grating. In combination, these spectrographs cover the full optical band and extend into the NIR at sufficient resolution for direct abundance analysis.  
A longslit mask of 1-arcsec width and 1-arcminute length is used as the target objects are typically not very extended, but even for extended objects, this slit provides adequate spatial coverage for accurate sky subtraction. The combination of sensitivity, resolution, and wavelength coverage of MODS permits high-quality optical spectroscopy for precise He/H measurements.

\subsubsection{Observing Strategy}

The airmasses of the observations varied depending on the time of observation, and, because the MODS spectrographs do not have atmospheric dispersion correctors, the celestial position angle of the slit was chosen to be the parallactic angle at mid-visit to minimize flux losses due to differential atmospheric  refraction \citep{fili1982} (MODS guide with a red filter with central wavelength of 660\,nm, so slit losses due to refraction would mostly impact the blue end of the spectra).  
We further restricted the observing windows for each target to airmass less than 1.5 (elevation $>42$\degr), and for those objects that pass within 5-degrees of zenith at the LBT, we avoid observing windows where the target would pass into elevations where the telescope instrument rotator must work near its maximum tracking rate (LBT is an alt-az telescope). 
Observations were taken when seeing was $<$1\farcs1, but we got acceptable spectra of compact targets in variable seeing conditions that exceeded this limit. 
Because observing time among the LBT members is allocated in week-long blocks that are queue scheduled, we were able to observe most of our planned targets over the course of the 3-year primary observing campaign with minimal weather losses.

In total, we obtained new, high-quality MODS optical spectra of 56 targets for a total of 70 hours (wall clock time) of LBT time (including overheads). In addition, we reduced and analyzed MODS spectra of 7 other targets taken from the LBT data archive, and we include our previous MODS observations of Leo~P \citep{aver2021} and AGC~198691 \citep{aver2022}.

Each new target was observed for 1-hour of open-shutter integration time with both MODS spectrographs identically configured, divided into three 20-minute exposures to allow clean removal of cosmic ray events from the raw images. 
A few targets could only be observed with one of the MODS spectrographs because the adaptive secondary mirror was unavailable on the other ``eye'' of the LBT. The goal was to obtain a sample where all targets were observed as homogeneously as possible with a ground-based telescope. Standard calibrations, flat fields, wavelength comparison lamps, and spectrophotometric standard stars, were obtained during each observing run. 

\subsubsection{Spectral Calibration}

Because the LBT has a small facility instrument suite that is mounted on the telescope essentially all of the time, except for during summer shutdown, calibrations are very stable, as instruments are rarely disturbed. All regular calibrations for all LBT observing programs are released immediately on the LBT Observatory Data Archive\footnote{\url{https://archive.lbto.org/}}, so we can examine all available calibration data, not just those obtained as part of our observing campaign, and get a long-term quantitative assessment of the overall quality of the flux calibration of MODS.  
Further, one of the authors (RWP) was the MODS project lead and had established a long-standing cross-partner calibration program whereby the same calibration procedures were followed for all internal instrumental calibrations (flat fields, wavelength calibration lamps, and biases) and flux standard star observations, so a high degree of calibration uniformity was maintained over many years before and after our campaign. 

For our observations in our previous two \Yp\ studies \citep{aver2021, aver2022}, a uniform, wavelength-independent flux calibration uncertainty of 2\% was assumed for all emission line measurements, based on the flux uncertainty quoted for the observations of standard stars \citep{oke1990}.  This was a very conservative assumption (clearly two emission lines which are near to each other in wavelength will have a much smaller minimum uncertainty in their relative fluxes).  
With the new observations from the LBT \Yp\ Project, we could revisit this assumption and replace it with a new, wavelength dependent calibration uncertainty.  

To derive MODS spectral response functions, we only used the best quality spectra of two HST CALSPEC \citep{bohl2014} primary flux calibrator DA white dwarf stars, G191B2b and GD153 \citep{bohlin2019}. 
Both stars had overlap between the observing semesters and were the highest priority flux stars in the overall observatory calibration program for MODS. Each star is observed with a special 5x60-arcsec  calibration slit mask to eliminate seeing- and refraction-dependent slit losses. We restricted attention to observations of these stars in the LBT Archive from 2021 through 2024 that covered our campaign, and we reduced all the spectra starting from the raw data using our version of the modsIDL pipeline\footnote{A description of the modsIDL pipeline can be found in\citet{berg2015}.} to make 1D extracted spectra, and then calculated response curves including atmospheric extinction using a custom Jupyter notebook. Poor-quality spectra taken through clouds or with other issues were culled from the final sample.  In total, our final sample consisted of 50 spectra of G191B2b and 19 spectra of GD153.

The historical response curves show the expected random changes in flux calibration due to weather; conditions at Mt. Graham are rarely photometric, but we see a well-defined upper envelop when it was photometric. Long-term trends seen in the LBT+MODS total throughput are understood as the accumulation of dust on the LBT primary mirrors and slow degradation of the exposed aluminum coatings.  The LBT primary mirrors are recoated on 2-year cycle and washed in between years, but, as our campaign started during the pandemic, this recoating schedule was interrupted. The adaptive secondary (AdSec) mirrors were generally used, but occasional technical downtime of one or other AdSec mirror required the observatory to swap in a rigid secondary mirror which has a slightly different aluminum coating. Various jumps in throughput seen are all aligned with observatory records of when the primary mirrors were washed or recoated and times when the rigid secondary was swapped in and out. 

The primary difference between MODS1 and MODS2 throughput during our campaign was that the aluminum coating on the adaptive secondary mirror on the left-side of the LBT, where MODS1 is mounted, had been steadily degrading for many years. The net effect of this degradation was an overall loss of throughput on the left side of LBT for all instruments of $\sim$0.5$-$0.6\,magnitudes. When the left adaptive secondary was recoated in summer 2023, the MODS1 throughput returned to near-historical values, but still $\sim$0.1 mag lower than the original MODS1 commissioning throughput due to the steady accumulation of dust on the internal spectrograph optics over 13 years of operation. MODS2, mounted on the right side of LBT, showed slow monotonic decline mostly due to the primary mirrors and some slow accumulation of dust on the internal optics, but a discontinuous change occurred after the recoating of the right-hand mirror in summer 2023, and the aluminum coating manifested a "blue haze" that degraded the throughput at blue wavelengths. We see this as a slight change in the slope of the MODS2 blue channel response functions before and after recoating. 

Overall, our retrospective analysis of MODS calibration star data spanning our campaign shows that the basic shape of the spectral response functions was as stable as can be expected for optical coatings exposed to the environment.  This allowed us to select a subset of the best calibration data from the entire span of the observing campaign for our calibrations. The calibrations for our sample are very well understood, something that is not possible for data sets collected from many disparate telescopes and instruments with no control over the calibration procedure or ability to perform a quantitative re-analysis of the calibration process.

\begin{figure}[t!]
\resizebox{\columnwidth}{!}{\includegraphics{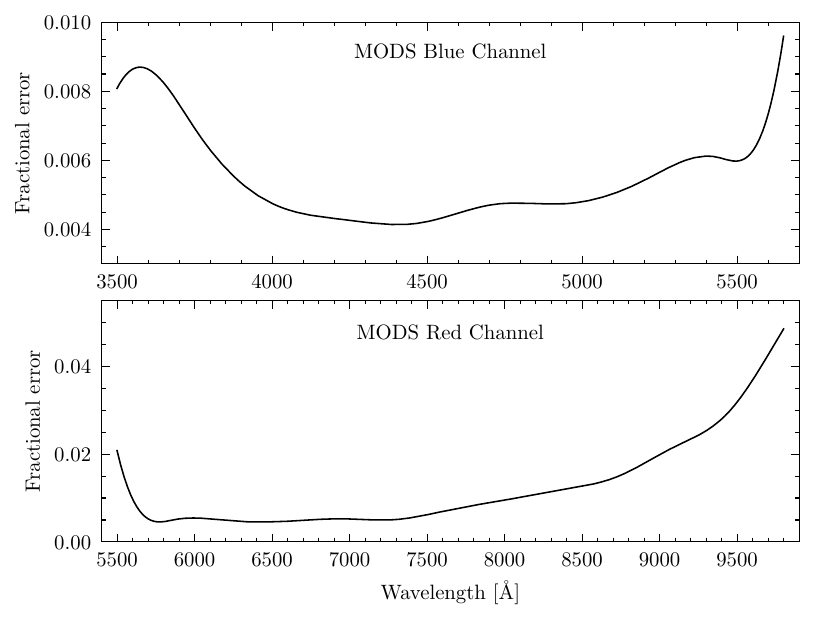}}
\caption{
Spectral flux calibration error functions for the MODS blue and red grating spectrograph
channels derived from archival and new observations of HST primary flux calibration stars. Curves show the fractional error in relative flux as a function of wavelength.
}
\label{fig:fluxCalErrs}
\end{figure}

Using the best calibration stars, we fit the response functions and, comparing with the cataloged primary flux calibrator data \citep{bohlin2019}, we compute estimates of the random errors in the spectrophotometry contributed by the relative spectral response calibration. The resulting error functions are shown in Figure\,\ref{fig:fluxCalErrs}.  The MODS blue channel relative flux calibration is typically good to $\pm$0.5\%, with the expected increase in error to $\pm1-2$\% at the extreme wavelength ends of the spectrum due to the combined fall-off in CCD detector quantum efficiency, grating blaze efficiency, and dichroic beam splitter transmission at the extreme blue/near-UV end of the spectrum, and into the dichroic cross-over at the red end.
The MODS red channel relative flux calibration is typically $\pm0.6$\% over most of the range, with the expected increase at the blue end from the dichroic cross-over, and the monotonic decline in CCD detector quantum efficiency and grating blaze efficiency at the red end. Unlike the blue channel, the red channel also has impacts due to telluric absorption starting at wavelengths long-ward of 6800\AA, which are difficult to correct for, leaving gaps in the response function that we interpolate across.  The interpolations are informed by a subset of standard stars for which we have been able to compute reasonable model telluric corrections for data taken at low airmass on nights with a low precipitable water-vapor (PWV) column density (at the time LBTO site weather telemetry did not provide consistent contemporaneous PWV measurements with the radiometer at the Hertz mm-wave telescope on the summit of Mt.\ Graham).  
The fidelity of these interpolations is high as changes in actual instrumental response are expected to be very smooth on smaller scales (due to the physics of CCDs and the stability of the optical coatings).

Details of the MODS data reduction and presentation and analysis of the final 1D optical spectra will be given in Paper II.

\subsection{Near-Infrared Spectroscopy}

\subsubsection{Observing Strategy}

Because the \ion{He}{1}\,$\lambda$10830\AA\ emission line provides a strong constraint on the nebular density and breaks the temperature-density degeneracy, we have used the LBT's LUCI (LBT Utility Camera in the Infrared) spectrographs \citep{seifert2003} to obtain a near-IR spectrum for most of the sample.  For a minority of the targets (11 targets), we have adopted values from the literature.  Over the  course of the program, both LUCI instruments were not always available, so some observations come from a single LUCI spectrograph, and some have observations from both instruments.

LUCI long-slit spectra were taken with the G200 grating (200\,lines\,mm$^{-1}$) in second order with the  zJ-band order-separation filter to obtain a spectral resolution of $\sim$2100 in the 0.9 to 1.35$\mu$m wavelength region that includes the \ion{He}{1}\,$\lambda$10830\AA\ and \ion{H}{1}\,P$\gamma$ emission lines over the range of redshifts of our targets.  We used the facility 1-arcsec wide long-slit mask (LS-1.00arcsec), which is 205-arcsec long, providing simultaneous measurement of the galaxy and  surrounding night sky. This is the same slit width used for the  MODS visible-light spectra. When LUCI1 and LUCI2 were both available they were configured identically. Observations were taken when visible-light seeing was $<$1\farcs1, although on a few occasions we obtained acceptable observations in variable seeing.  
LUCI spectra were prioritized by the outcome of earlier MODS observing during our blocks of runs, but we never observed with both instruments on the same night.  In total, we obtained high-quality LUCI near-IR spectra of 48 targets new for a total of 72 hours (wall clock time) of LBT time.  We include our archival LUCI spectra of Leo~P \citep{aver2021} in our analysis for a total of 49 LUCI observations.

Each target was observed for 1-hour total integration time acquired as a sequence of six (6) 600-second integrations in which we offset between two positions separated by 10\arcsec\ along the slit in an ABBAAB dither pattern.  In a small number of galaxies, nearby structures along the slit required us to use a 20\arcsec\ dither spacing. This dither pattern allows for very clean sky subtraction in a spectral region which is crowded with many bright telluric OH airglow emission lines.  All targets were observed at low airmass, and in this spectral range differential atmospheric refraction is negligible, unlike the case with MODS.

The wavelength calibration was determined using terrestrial OH airglow emission lines that appear in the longslit spectra with the wavelengths tabulated by \citet{rousselot2000}, and we used quartz lamp spectral flats and darks obtained on or close to the nights of LUCI observation. Like the MODS, the LUCIs are mounted on the telescope full-time except for summer shutdowns, and historically have had very stable internal detector calibrations. Bad pixel maps for the LUCIs' H2RG HgCdTe arrays are historically very stable, our targets are faint enough that we did not have problems with persistent image on the sensors, and we were careful when observing tellurics before science targets to not saturate the arrays.

\subsubsection{Spectral Calibration}

Spectral calibration for near-IR observing requires a different approach than the optical, relying traditionally on observations of A0V and A1V spectral type telluric stars selected using the Gemini Observatory Telluric Standard Search tool\footnote{\url{https://www.gemini.edu/observing/resources/near-ir-resources/spectroscopy/telluric-standard-search}}. We selected telluric stars matched to each target such that they would be observed at the same airmass as the target either before or after the science observations.  
The telluric star spectra provide a correction for atmospheric absorption in the zJ band, and a baseline relative flux calibration around the \ion{He}{1} line.  However, unlike with the MODS, the standard practice with the LUCIs is to not do a detailed end-to-end response calibration.  
Instead, we rely on the fact the relative fluxes of lines of interest are closely spaced and the measured spectral response functions for LUCI are very well behaved on such small wavelength scales. However, we made observations of the same HST primary calibrators we used with MODS on two observing runs and verified our assumptions are correct.

Details of the LUCI data reduction and presentation and analysis of the final 1D near-IR spectra will be given in Paper III.



\section{New H and He Emissivities}

\subsection{H emissivities}

Our previous work used the ``Case B'' H emissivities presented in \citet{hummer1987}.
These had been updated by \citet{storey2015}.  \citet{hsyu2020} reported a new set of H emissivities from P.\ Storey (2018, private communication).
These emissivities extend the \citet{storey2015} hydrogen emissivities down to log$_{10}(n_{\rm e}/\rm cm^{-3}$)\,=\,0, and are calculated on a fine grid up to log$_{10}(n_{\rm e}/\rm cm^{-3}$)\,=\,5 at log$_{10}(n_{\rm e}/\rm cm^{-3}$)\,=\,1 intervals.
The grid used by \citet{hsyu2020} stopped at the n $=$ 10 level, and was therefore not suitable for our purpose.  
P.\ Storey (2023, private communication) provided us with an even finer grid extended to the n $=$ 15 level.

We use bilinear interpolation within this temperature and density grid.
We assume no error in the emissivity value as this is difficult to quantify, and we assume that the uncertainties on the H emissivities are sub-dominant to the other uncertainties in our procedure. 
This assumption is supported by the work of \citet{hsyu2020} who compared the new H$\beta$ emissivity with that of R.\ L.\ Porter, as presented in \citet{aver2010}.  They found differences in the range of 0.10 to 0.55\%.  \citet{hsyu2020} also compared the new \citet{storey2015} emissivities for the three strongest Balmer line ratios to the \citet{aver2010} fit \citep[based upon][]{hummer1987} and found differences in the range from 0.10 to 0.20\%.  These very small differences between the most recent calculations support our assumption that the uncertainties on the H emissivities are negligible.
The details of these new H emissivities are discussed in Paper IV.

\subsection{He Emissivities}

Our previous work used the ``Case B'' He emissivities from \citet{porter2012, porter2013} calculated on a finer grid, as reported in \citet{aver2013}. 
Motivated by their discovery that the He recombination spectrum could be significantly improved \citep{DelZanna2020}, \citet{DelZanna2022} have produced new He emissivities, and, unlike the case for H, found differences in excess of 1\% when compared to previous studies.  They provide a comprehensive overview of and comparison with previous studies.
The biggest change relevant for our purpose is found in the $\lambda$6678 line, which shows good agreement at low temperatures, but increasing discrepancies as the density and temperature increase.
We adopt these new He emissivities, and, although the case is not as secure as for the H emissivities, we assume no uncertainties in the He emissivities, given the lack of quantified estimates.
The details of these new He emissivities are discussed in Paper IV.



\section{LBT $Y_{\rm p}$ Project 1D Spectral Database}

Because of their potential interest to the wider community, the publication of the primary results of this project will be accompanied by release of our 1D spectral data products in the form of machine-readable tables of our measured spectral lines fluxes in optical and infrared, and machine-readable ASCII tables of the extracted, wavelength- and flux-calibrated 1D spectra.

The near-infrared 1D spectral database will consist of the measured spectral line fluxes and 1D calibration spectra will be released as part of Paper III.  The 1D merged LUCI spectra will be 3-column ASCII tables with wavelengths, relative fluxes, and flux errors.

The optical 1D spectral database will be released in two parts. The tables of hydrogen and helium emission line fluxes and uncertainties and derived nebular properties (O/H, temperature, density, etc.) used in the analysis in Paper IV will be released as part of Paper~II.  The full measured spectral line database (all species, not just the H and He lines), and the 1D wavelength- and flux calibrated MODS spectra will be released as part of Paper~VI.  The format of the 1D merged MODS spectra will be in the same 3-column ASCII format as the LUCI near-infrared spectra.

\section{Summary}
\subsection{Overall Project Goal}

A simple back of the envelope calculation tells us that the helium mass fraction is $\sim 25$\% from SBBN. 
That the helium abundance is observed to be around 25\% played a big role in supporting the big bang theory over competing cosmological theories, as stars alone are not capable of producing the requisite helium abundance. However, to fully test big bang cosmology and Standard Model physics, high-precision data are needed. Historically the helium abundance in big bang nucleosynthesis depends on three quantities: the baryon density, the neutron mean-life, and the number of relativistic species present at the time of nucleosynthesis \citep{Olive:1980bu}. 
The latter can be characterized by the number of neutrino flavors. Precision on the baryon density is achieved from CMB measurements of the anisotropy spectrum (first by WMAP \citep{komatsu2014} and later by {\it Planck} \citep{aghanim2020}). The neutron mean-life is also known to high precision ($\sim$0.05\%) though there is currently a tension between different methods of determining the neutron lifetime. Its impact on BBN was recently reviewed in \citet{Yeh:2023nve}. In the Standard Model, $N_\nu = 3$, and measured as $N_\nu = 2.9963 \pm 0.0074$ from the width of the $Z$ gauge boson \citep{Janot:2019oyi}. This leads to the SBBN prediction $Y_P = 0.2467 \pm 0.0002$ \citep{Yeh:2022heq,Yeh:2023nve}, with better than $0.1$\% accuracy. Ultimately, we would like a comparable observational uncertainty to fully test the theory. In comparison, the current accuracy of the measured deuterium abundance is about 1\%, though the theoretical uncertainty is much larger in this case. 

Decades ago, helium abundance determinations were based on averaging the independently inferred abundances from three helium emission lines. The quoted uncertainties were typically of order 1\%. These however ignored a whole host of systematic uncertainties. \citet{olive2001,olive2004} showed that there exist severe degeneracies in assumed physical parameters that, until resolved, would not permit a helium abundance determination to better that $\sim 5$\%. With a more rigorous approach to extracting a helium abundance, it was possible to obtain an uncertainty as low as $\sim 1.3$\% \citep{aver2022}. This determination still relied on a regression of the data with respect to metallicity with the primordial value resulting from the intercept of that regression. This procedure itself carries its own uncertainties, as we can not be sure that helium and oxygen track each other linearly. After all, there are different processes in different stars which produce these two elements. 

Our goal in this series of papers is to produce an XMP sample of galaxies with high emission line fluxes. With a sufficiently large sample size of XMP galaxies, we could avoid relying on a regression (or assumed functional dependence) between helium and oxygen. We believe that we have made an important step towards this goal as detailed in this series of papers.

\subsection{Improvements to Our Methodology}

Our LBT \Yp\ Project methodology is built on the analyses reported in 
\citet{aver2015, aver2021, aver2022}. As described here, we have identified required characteristics of candidate targets (low O/H, high H$\beta$ flux, high H$\beta$ equivalent width) in order to produce high-quality, useful measurements.  We have optimized data reduction and analyses processes to produce high fidelity measurements with reliable uncertainties.  We use the most modern H and He emissivities calculated on very fine grids.  We use the most modern radiative transfer corrections following \citet{kurichin2025}. Other more subtle improvements in the analysis are describe in Paper~IV.
In summary, in addition to gathering the desired high quality LBT observations, we have sought to improve every step of the data reduction and analysis chain.

\subsection{Papers in this Series}

In Paper II \citep{Rogers2026}, we describe the reduction of the LBT/MODS optical observations. This includes a new treatment of the emission line profiles in MODS, allowing for better constraints on emission line fluxes of blended features such as \ion{He}{1} $\lambda$3889 and H8. The optical line fluxes are used to derive the physical conditions of the gas and measure the total O/H abundance. The direct T$_e$[\ion{O}{3}] in the interstellar medium is used as a weak prior on T$_e$(\ion{He}{1}) in the MCMC analysis, but also provides an opportunity to examine scaling relations between T$_e$ from other ions (e.g., S$^{2+}$ and O$^{+}$). The O/H abundances, which are traditionally required for the extrapolation of $\Delta$Y/$\Delta$O, have typical uncertainties of just 4\%.

In Paper III \citep{Weller2026}, we describe the reduction of the LBT/LUCI observations.  These observations include the critical \ion{He}{1}\,$\lambda$10830 line, which provides very strong constraints on the density. Improvements and optimizations upon the standard LUCI reduction pipeline are described, enabling robust, high quality measurements of the near-IR emission lines.

Paper IV \citep{Aver2026} presents the derivations of the helium abundances of the individual objects, the screening of the sample for reliability, and the subsequent derivation of a new value of \Yp.  The entire sample consists of 41 objects all with O/H $\le 14.5 \times 10^{-5}$, and 15 of these objects have metallicities of O/H $\le 4 \times 10^{-5}$.  This low-metallicity sample of 15 objects is in the regime where evolutionary effects are minimal, allowing a derivation of \Yp\ from the weighted mean, free from the assumption of a linear relationship between Y and O/H. 
Full details of the analysis of the entire data set are provided in Paper IV. 

Implications of the new result for $Y_{\rm p}$ are presented in Paper V \citep{Yeh2026}. In particular, the new value of $Y_{\rm p}$ and its uncertainty are used to determine the energy density in radiation at the time of BBN. This is then used to constrain the number of relativistic degrees of freedom present at that time. This result is presented as a limit on the number of neutrino flavors and can be adapted to constrain a variety of physics models extending beyond the Standard Model \citep{Cyburt2005,Yeh:2022heq}.

The spectral database assembled as part of the LBT \Yp\ Project can be tapped for other science cases. In Paper VI (N.\ Rogers et al., in prep), we explore the chemical abundances of nine elements in the metal-poor interstellar medium. This represents one of the largest homogeneous databases of high-quality multi-element abundances in low-$Z$ nebulae, which is required to interpret emerging abundance trends at high-$z$ acquired with JWST.



\begin{acknowledgements}
This work was supported by funds provided by NSF Collaborative Research Grants AST-2205817 to RWP, AST-2205864 to EDS, and AST-2205958 to EA. 
Our team workshop at OSU in July 2024 was sponsored in part by OSU's Center for Cosmology and AstroParticle Physics (CCAPP). The work of KAO was supported in part by DOE grant DE-SC0011842 at the University of Minnesota. 
TRIUMF receives federal funding via a contribution agreement with the National Research Council of Canada.

This work is based on observations made with the Large Binocular Telescope. The LBT is an international collaboration among institutions in the United States, Italy and Germany. LBT Corporation Members are: The University of Arizona on behalf of the Arizona Board of Regents; Istituto Nazionale di Astrofisica, Italy; LBT Beteiligungsgesellschaft, Germany, representing the Max-Planck Society, The Leibniz Institute for Astrophysics Potsdam, and Heidelberg University; The Ohio State University, and The Research Corporation, on behalf of The University of Notre Dame, University of Minnesota and University of Virginia. Observations have benefited from the use of ALTA Center (alta.arcetri.inaf.it) forecasts performed with the Astro-Meso-Nh model. Initialization data of the ALTA automatic forecast system come from the General Circulation Model (HRES) of the European Centre for Medium Range Weather Forecasts.

This research used the facilities of the Italian Center for Astronomical Archive (IA2) operated by INAF at the Astronomical Observatory of Trieste.

All LBT observations for this project were executed remotely, beginning during the global COVID-19 pandemic when we all learned together how to operate the LBT from basements, spare bedrooms, and home offices. We are most grateful for the tireless efforts of the Mount Graham and Tucson technical and observing support staff of the Large Binocular Telescope Observatory who were critical to making it all work smoothly, keeping telescopes and instruments operating in top form under trying circumstances. The 130 hours of high-quality, consistent spectrophotometric data acquired for this project would not have been possible without them.

EDS, KAO, EA, DAB, NSJR, and JHM would like to acknowledge and thank Stanley Hubbard for his generous gift to
the University of Minnesota that allowed the University to become a member of the LBT collaboration.

\end{acknowledgements}

\appendix
\vspace{-0.5cm}

\section{Methodological Differences Between Aver et al.\ (2015) and Hsyu et al.\ (2020)}

\citet{hsyu2020} independently coded the methodology of \citet{aver2015} for their analysis.  However, beyond any small differences in atomic data or number of emission lines used, there are some subtle but potentially important differences in the methodology.  Specifically, \citet{aver2015} use a $\chi$$^2$ minimization to determine the best-fit (maximum likelihood) values for the model parameters and then follow with an MCMC analysis to determine the uncertainties on those values.  The MCMC analysis is critical because it reveals degeneracies between various parameters \citep{aver2011}.  While 
\citet{hsyu2020} follow the \citet{aver2015} MCMC, they adopt final values for the parameters from the median values of the MCMC distribution.  In typical cases, the minimization best fits and the median values from the MCMC analyses will be quite similar.  However, in the cases where parameters are restricted to positive values (e.g., density, underlying stellar absorption, optical depth), and when the maximum likelihood values are small, the distribution of Markov Chain values will be truncated to positive values, with an asymmetric tail extending to higher values.  This results in a median value significantly higher than the best-fit, maximum likelihood value.  Furthermore, in the cases of parameters with maximum likelihood values approaching the zero lower bound, using the median value results in a lower limit for the parameter's uncertainty range that does not even encompass the maximum likelihood value. The impact of this bias is difficult to predict, but there will be departures from the best-fit values.  

A second difference is in how objects ``qualify'' for inclusion in the determination of $Y_{\rm p}$.  \citet{aver2012} introduced (within He abundance analyses) the concept of using the minimized $\chi^2$ value (i.e., the best model fit to the observations) as a criterion for objects to qualify for use in the determination of $Y_{\rm p}$.  This provides a very direct and statistically rigorous way to assess the reliability of the results. \citet{hsyu2020} use a different criterion.  In order for objects to qualify in their analysis, they require that the measured and model-predicted values for all of their emission lines agree to within two $\sigma$ (their Sample~1) or all but one line agree to within two $\sigma$ (their Sample~2).  Clearly, if their median parameter values are near their maximum likelihood values (as would be hoped), these both represent a lower threshold and allow a larger number of objects to be used in their $Y_{\rm p}$ analysis.  It is not clear what the impact of this choice is on the value of $Y_{\rm p}$, but it is likely that the uncertainty is reduced by including more objects in the analysis that likely would not have qualified using a $\chi^2$ criterion.

A third important difference is the inhomogeneity of the observations in their sample.  The importance of observing the \10830\ emission line has been emphasized and they obtain \10830\ observations for the 3 galaxies from their PHLEK sample and the 13 galaxies from the HeBCD sample have existing 10830 observations; however, they do not have \10830\ observations for the 38 galaxies in their SDSS sample, the majority of the 54 systems in their Sample~1.  We are not sure what the potential systematic effects are of not limiting the sample to systems with  \10830\ observations, beyond larger per object uncertainties on the derived He/H abundances

\citet{matsumoto2022} and \citet{yanagisawa2025} adopt the \citet{hsyu2020} methodology, so the same differences as detailed above apply, with one revision.  Motivated by the concerns above, \citet{matsumoto2022} allows for unphysical, negative parameter values when analyzing the Subaru observations in their sample \citep[though not for the literature values they take from][]{hsyu2020}.  Our approach is to retain the positive restriction, but use the maximum likelihood value, rather than the median value.  Please see Section \ref{Sec:RecentResults} for further details on the samples, analysis, and results in \citet{matsumoto2022} and \citet{yanagisawa2025}.

\facilities{LBT (MODS), LBT (LUCI)}
\software{
\texttt{astropy} \citep{astr2013, astr2018, astr2022},
\texttt{jupyter} \citep{kluy2016},
\texttt{modsIDL} \citep{crox2019},
\texttt{modsCCDRed} \citep{pogg2019},
\texttt{PypeIt} \citep{pypeit:joss_pub, pypeit:zenodo}
\texttt{PyNeb} \citep{luri2012,luri2015L},
\texttt{numpy} \citep{harr2020},
\texttt{matplotlib} \citep{hunt2007}
}


\newpage
\clearpage

\bibliographystyle{aasjournalv7}
\bibliography{yp_total_bib}

\clearpage

\end{document}

%% file: ObservationsTable.tex
\begin{deluxetable}{lcccrrr}
\label{table:Observations}
\centering
\tablecaption{LBT Observations for the LBT $Y_{\rm p}$ Project
\label{t:observations}}
\tabletypesize{\scriptsize}
\tablewidth{\textwidth}
\tablehead{ &   R.A.    & Decl. &   &   \multicolumn{2}{c}{Int. Time (min)} &   \\
Object  & (hh:mm:ss)    & (deg:mm:ss)   & z &   MODS    & LUCI  &   Other Names(s)/Notes
}
\startdata
AGC 198691	&	09:43:32.38	&	+33:26:57.9	&	0.0017	&	120	&	Literature	&	Leoncino, Previously Reported	\\
DDO 68	&	09:56:46.80	&	+28:50:10.9	&	0.0017	&	160	&	160	&	UGC 5340	\\
DESI J092331.28+645111.3	&	09:23:31.27	&	+64:51:11.3	&	0.0054	&	120	&	\nodata	&	Not Viable	\\
HS 0029+1748	&	00:32:03.11	&	+18:04:46.1	&	0.0071	&	60	&	60	&		\\
HS 0122+0743 	&	01:25:34.20	&	+07:59:24.0	&	0.0098	&	60	&	60	&	UGC 993	\\
HS 0134+3415	&	01:37:13.80	&	+34:31:12.0	&	0.0195	&	120	&	60	&		\\
HS 0811+4913	&	08:14:47.53	&	+49:04:00.7	&	0.0018	&	120	&	60	&	SHOC 193b	\\
HS 0837+4717	&	08:40:29.91	&	+47:07:10.2	&	0.0420	&	60	&	Literature	&	SHOC 220	\\
HS 1028+3843	&	10:31:51.63	&	+38:28:08.4	&	0.0295	&	60	&	120	&		\\
HS 1222+3741	&	12:24:36.72	&	+37:24:36.6	&	0.0404	&	120	&	120	&		\\
HS 1353+4706	&	13:55:25.66	&	+46:51:51.3	&	0.0281	&	120	&	120	&		\\
HS 1442+4250	&	14:44:11.48	&	+42:37:35.9	&	0.0021	&	60	&	120	&	UGC 9497	\\
HS 2236+1344	&	22:38:31.12	&	+14:00:29.8	&	0.0206	&	80	&	60	&		\\
HSC J2314+0154	&	23:14:37.55	&	+01:54:14.3	&	0.0327	&	80	&	\nodata	&		\\
I Zw 18 NW	&	09:34:02.17	&	+55:14:27.4	&	0.0025	&	120	&	\nodata	&	Archival MODS	\\
I Zw 18 SE	&	09:34:01.97	&	+55:14:28.2	&	0.0025	&	120	&	90	&		\\
J0118+3512	&	01:18:40.00	&	+35:12:57.0	&	0.0154	&	120	&	Literature	&		\\
J0519+0007	&	05:19:02.64	&	+00:07:23.0	&	0.0444	&	120	&	Literature	&		\\
J2213+1722 	&	22:13:48.45	&	+17:22:35.6	&	0.0048	&	120	&	Literature	&		\\
KKH 46	&	09:08:36.54	&	+05:17:27.0	&	0.0021	&	80	&	\nodata	&	Not Viable	\\
KUG 0743+513	&	07:47:33.18	&	+51:11:24.8	&	0.0014	&	120	&	120	&		\\
KUG 1138+327	&	11:41:07.49	&	+32:25:37.2	&	0.0060	&	60	&	120	&	LEDA 36252, Archival MODS	\\
LEDA 101527	&	15:09:34.17	&	+37:31:46.1	&	0.0326	&	160	&	80	&		\\
LEDA 2790884	&	08:25:55.52	&	+35:32:32.0	&	0.0020	&	120	&	Literature	&		\\
Leo P	&	10:21:45.10	&	+18:05:17.1	&	0.0009	&	45	&	120	&	AGC 208583, Previously Reported	\\
Mrk 5                 	&	06:42:15.26	&	+75:37:30.4	&	0.0016	&	120	&	60	&	UGCA 130	\\
Mrk 36	&	11:04:58.54	&	+29:08:15.7	&	0.0022	&	60	&	Literature	&	UGCA 225, Archival MODS	\\
Mrk 71	&	07:28:42.80	&	+69:11:21.0	&	0.0004	&	60	&	60	&	NGC 2363A	\\
SBS 0335-052E	&	03:37:44.06	&	-05:02:40.2	&	0.0135	&	120	&	120	&		\\
SBS 0926+606	&	09:30:09.05	&	+60:28:05.4	&	0.0138	&	120	&	60	&	LEDA 26955	\\
SBS 0940+544	&	09:44:16.61	&	+54:11:34.3	&	0.0054	&	60	&	Literature	&		\\
SBS 0946+558          	&	09:49:30.30	&	+55:34:47.0	&	0.0052	&	120	&	60	&	Mrk 22	\\
SBS 0948+532          	&	09:51:31.77	&	+52:59:36.0	&	0.0462	&	120	&	60	&	LEDA 28398	\\
SBS 1030+583	&	10:34:10.15	&	+58:03:49.1	&	0.0075	&	60	&	Literature	&	Mrk 1434, Archival MODS	\\
SBS 1135+581	&	11:38:35.68	&	+57:52:27.3	&	0.0032	&	120	&	Literature	&	Mrk 1450	\\
SBS 1152+579	&	11:55:28.34	&	+57:39:52.0	&	0.0173	&	120	&	Literature	&	Mrk 193	\\
SBS 1159+545	&	12:02:02.37	&	+54:15:49.6	&	0.0120	&	120	&	120	&	LEDA 2815986	\\
SBS 1211+540          	&	12:14:02.48	&	+53:45:17.4	&	0.0031	&	60	&	120	&		\\
SBS 1249+493          	&	12:51:52.50	&	+49:03:27.7	&	0.0243	&	120	&	120	&		\\
SBS 1331+493	&	13:33:23.84	&	+49:06:12.8	&	0.0020	&	120	&	120	&		\\
SBS 1415+437	&	14:17:01.41	&	+43:30:05.5	&	0.0021	&	60	&	120	&		\\
SBS 1420+544          	&	14:22:38.85	&	+54:14:09.2	&	0.0206	&	120	&	120	&		\\
SBS 1437+370	&	14:39:05.47	&	+36:48:21.0	&	0.0019	&	60	&	Literature	&	Mrk 475	\\
SDSS J013352.55+134209.5	&	01:33:52.56	&	+13:42:09.4	&	0.0088	&	120	&	120	&		\\
SDSS J080758.0+341439.3	&	08:07:58.00	&	+34:14:39.3	&	0.0225	&	120	&	60	&		\\
SDSS J114827.33+254611.7	&	11:48:27.34	&	+25:46:11.8	&	0.0451	&	120	&	120	&	Archival MODS	\\
SDSS J210455.31-003522.2	&	21:04:55.31	&	-00:35:22.2	&	0.0048	&	60	&	100	&		\\
SHOC 113	&	02:18:52.91	&	-09:12:18.8	&	0.0127	&	120	&	120	&	Not Viable	\\
SHOC 133	&	02:40:52.21	&	-08:28:27.4	&	0.0822	&	60	&	\nodata	&		\\
SHOC 357	&	12:01:22.31	&	+02:11:08.4	&	0.0033	&	\nodata	&	120	&	LEDA 3120683	\\
UGC 4483	&	08:37:03.08	&	+69:46:50.3	&	0.0005	&	60	&	120	&		\\
UGC 5541	&	10:16:53.12	&	+58:23:40.3	&	0.0078	&	120	&	120	&		\\
UGC 6456	&	11:28:00.97	&	+78:59:28.3	&	-0.0003	&	120	&	120	&		\\
UM 133                	&	01:44:41.28	&	+04:53:26.0	&	0.0054	&	120	&	60	&		\\
UM 161	&	23:27:43.69	&	-02:00:55.9	&	0.0181	&	60	&	120	&		\\
UM 420	&	02:20:54.51	&	+00:33:23.6	&	0.0584	&	120	&	60	&	SBS 0218+003	\\
UM 461	&	11:51:33.35	&	-02:22:21.9	&	0.0035	&	120	&	120	&		\\
UM 570	&	13:23:47.46	&	-01:32:52.0	&	0.0225	&	120	&	120	&	SHOC 424, Archival MODS	\\
VCC 1744	&	12:38:06.89	&	+10:09:56.0	&	0.0038	&	120	&	120	&	LEDA 42204	\\
WISEA J085115.60+584055.7	&	08:51:15.65	&	+58:40:55.0	&	0.0919	&	120	&	60	&		\\
WISEA J104457.84+035312.9	&	10:44:57.79	&	+03:53:13.2	&	0.0130	&	90	&	120	&	Archival MODS	\\
WISEA J113623.81+470928.9	&	11:36:23.82	&	+47:09:29.1	&	0.0102	&	120	&	120	&		\\
WISEA J120503.54+455151.0	&	12:05:03.55	&	+45:51:50.9	&	0.0654	&	120	&	\nodata	&		\\
WISEA J132723.31+402203.5	&	13:27:23.29	&	+40:22:04.2	&	0.0105	&	160	&	120	&		\\
WISEA J133126.86+415148.5	&	13:31:26.91	&	+41:51:48.3	&	0.0117	&	120	&	120	&		\\
WISEA J231048.92-021105.9	&	23:10:48.84	&	-02:11:05.7	&	0.0125	&	120	&	120	&		\\
\enddata
\end{deluxetable}